\documentclass[12pt, prd, showpacs]{revtex4}
\usepackage{amssymb}
\usepackage{amsmath}

\input{tcilatex}

\begin{document}

\title{Acceleration of particles by black holes: kinematic explanation}
\author{O. B. Zaslavskii}
\affiliation{Department of Physics and Technology, Kharkov V.N. Karazin National
University, 4 Svoboda Square, Kharkov, 61077, Ukraine}
\email{zaslav@ukr.net }

\begin{abstract}
A new simple and general explanation of the effect of acceleration of
particles by black holes to infinite energies in the centre of mass frame is
suggested. It is based on kinematics of particles moving near the horizon.
This effect arises when particles of two kinds collide near the horizon. For
massive particles, the first kind represents a particle with the generic
energy and angular momentum (I call them "usual"). Near the horizon, such a
particle has a velocity almost \ equal to that of light in the frame that
corotates with a black hole (the frame is static if a black hole is static).
The second kind (called "critical") consists of particles with the velocity $%
v<c$ near the horizon due to special relationship between the energy and
angular momentum (or charge). As a result, the relative velocity approaches
the speed of light $c$, the Lorentz factor grows unbound. This explanation
applies both to generic rotating black holes and charged ones (even for
radial motion of particles). If one of colliding particles is massless
(photon), the critical particle is distinguished by the fact that its
frequency is finite near the horizon. The existence (or absence) of the
effect is determined depending on competition of two factors - gravitational
blue shift for a photon propagating towards a black hole and the Doppler
effect due to transformation from the locally nonrotating frame to a
comoving one. Classification of all possible types of collisions is
suggested depending on whether massive or massless particle is critical or
usual.
\end{abstract}

\maketitle

\section{Introduction}

Recently, an interesting effect was discovered. It turned out that two
particle can collide near the horizon of the Kerr black hole in such a way
that the energy in the centre of mass frame grows unbound \cite{ban} (we
will call it the BSW effect). This provoked a series of consequent papers 
\cite{berti} - \cite{gs} in which details of collision were studied, the
effect has been found in different and more general space-times, etc. The
key role in the BSW effect is played by the fact that one of two colliding
particles should be "critical". By definition, this means that its energy $E$
and angular momentum $L$ are connected by the relationship $E-\omega _{H}L=0$
where $\omega _{H}$ is the angular velocity of \ black hole. (Otherwise, I
call a particle "usual"). For the Kerr metric, this was observed in \cite%
{ban} and traced in detailed in subsequent papers \cite{gp4}, \cite{inter},
including even nonequatorial motion \cite{kerr}. For a generic rotating
axially-symmetric dirty black hole (surrounded by matter) this was found in 
\cite{prd}. In Ref. In \cite{cqg} the most general geometric explanation was
suggested that relies on the relative orientation of the particle's timelike
four-velocity and generator of a black hole horizon.

The aim of the present work is to give an alternative, purely kinematic
explanation of the BSW effect with the emphasis on the role of critical
particles in the terms of particles' three-velocities. To the best of my
knowledge, this was not done yet. In Ref. \cite{ban} the remark has been
made by passing that the effect is connected with a crucial difference
between kinematics of "usual" and "critical" particles. In the first case, a
particle hits the horizon of a rotating black hole perpendicularly, in the
second one it does it at some incident angle. This important observation
does not give, however, \ the full explanation of the phenomenon. For
example, the effect exists even for pure radial motion of charged particles
in the Reissner-Nordstr\"{o}m space-time \cite{jl} when, obviously, all
particle approach horizon perpendicularly. From another hand, two different
critical particles can collide at some nonzero angle if they have different
momenta. However, this does not produce the infinite energy in the centre of
mass frame. Thus, an interesting property mentioned in \cite{ban} is neither
necessary nor sufficient for the explanation of the effect under discussion.

Below, in Sec. II\ - V I show that for the collision of massive particles,
the crucial point is whether or not a particle has near the horizon the
velocity approaching the speed of light. Only collision of particles of both
different kinds produce the effect. The frame in which the velocity under
discussion is measured is either the static one (for charged static black
holes) or the frame of the zero angular momentum observer (ZAMO) \cite{aj72}%
. Then, as we will see, the essence of the effect can be understood in the
terms of special relativity in combination with general consequences of
geodesic motion near the horizon.

In Sec. VI, I consider separately the case of collision between massive
particles and massive and massless ones since explanation for the BSW effect
and the definition and a role of critical particle is somewhat different in
both cases. Collisions of this type were considered in \cite{kerr} but for
the concrete case of the Kerr metric only.

In what follows we assume (as usual for the BSW effect) that both colliding
particles are ingoing. (The case when one of two particles is outgoing is
more simple and always leads to infinite energies simply due to blue shift
near the horizon \cite{pir}.) The effect under consideration is interesting
both from the theoretical viewpoint (since it gives hope to probe Planck
physics near the black hole horizon) and the astrophysical one (since
collisions with divergent energies can, in principle, leave their imprint on
the emergent flux of particles escaping from a black hole \cite{flux}, \cite%
{aw}).

\section{Basic equations}

Let us consider the space-time of a rotating black hole described by the
metric

\begin{equation}
ds^{2}=-N^{2}dt^{2}+g_{\phi \phi }(d\phi -\omega dt)^{2}+dl^{2}+g_{zz}dz^{2}.
\label{m}
\end{equation}%
Here, the metric coefficients do not depend on $t$ and $\phi $. On the
horizon $N=0$. Alternatively, one can use coordinates $\theta $ and $r$,
similar to Boyer--Lindquist ones for the Kerr metric, instead of $l$ and $z$%
. In (\ref{m}) we assume that the metric coefficients are even functions of $%
z$, so the equatorial plane $\theta =\frac{\pi }{2}$ ($z=0$) is a symmetry
one. The explicit form of the metric coefficients is not specified, so
consideration applies to "dirty" black holes surrounded by matter in
equilibrium with the horizon.

We consider the geodesic motion of massive particles in the equatorial plane 
$\theta =\frac{\pi }{2}$. The equations of motion have the form%
\begin{equation}
\dot{t}=u^{0}=\frac{E-\omega L}{N^{2}}\text{,}  \label{0}
\end{equation}%
\begin{equation}
\dot{\phi}=\frac{L}{g_{\phi \phi }}+\frac{\omega (E-\omega L)}{N^{2}},
\end{equation}%
\begin{equation}
\dot{l}^{2}=\frac{(E-\omega L)^{2}}{N^{2}}-1-\frac{L^{2}}{g_{\phi \phi }}.
\label{1}
\end{equation}%
where $E=-u_{0}$ and $L=u_{\phi }$ are conserved energy and angular momentum
per unit mass, $u^{\mu }$ is the four-velocity. In the present paper I use
units in which the gravitational constant $G=1$ and the speed of light $c=1$.

We assume that $\dot{t}>0$, so that $E-\omega L>0$ (motion forward in time),
except, possibly on the horizon where we admit the equality $E-\omega _{H}L=0
$ (subscript "H" denotes quantities calculated on the horizon). By
definition, if $E-\omega _{H}L>0$ a particle is "usual" and if $E-\omega
_{H}L=0$ it is "critical".

In what follows we will use the tetrad basis. Denoting coordinates $x^{\mu }$
as \ $x^{0}=t,x^{1}=l$, $x^{2}=z$, $x^{3}=\phi $, we choose the tetrad
vectors $h_{(a)\mu }$ in the following way:%
\begin{equation}
h_{(0)\mu }=-N(1,0,0,0)\text{, }  \label{h0}
\end{equation}%
\begin{equation}
h_{(1)\mu }=(0,1,0,0),
\end{equation}%
\begin{equation}
h_{(2)\mu }=\sqrt{g_{zz}}(0,0,0,1),
\end{equation}%
\begin{equation}
h_{(3)\mu }=\sqrt{g_{\phi \phi }}(-\omega ,0,0,1)  \label{h3}
\end{equation}%
If such a tetrad is attached to an observer moving in the metric (\ref{m}),
it has meaning of zero angular momentum observer (ZAMO) \cite{aj72}. They
are "rotate with the geometry" in the sense that $\frac{d\phi }{dt}\equiv
\omega $ for them. The advantage of using the tetrad components consists in
that one can use the formulas of special relativity in the flat space-time
tangent to any given point.

Then, we can introduce the three-velocity according to%
\begin{equation}
v^{(i)}=v_{(i)}=\frac{u^{\mu }h_{\mu (i)}}{-u^{\mu }h_{\mu (0)}}.  \label{vi}
\end{equation}

One can check that

\begin{equation}
-u_{\mu }h_{(0)}^{\mu }=\frac{E-\omega L}{N},
\end{equation}%
\begin{equation}
u_{\mu }h_{(3)}^{\mu }=\frac{L}{\sqrt{g_{\phi \phi }}}\text{.}
\end{equation}

From equations of motion (\ref{0}) - (\ref{1})\ and formulas for tetrad
components, we obtain%
\begin{equation}
v^{(3)}=\frac{LN}{\sqrt{g_{\phi \phi }}(E-\omega L)},  \label{v3}
\end{equation}%
\begin{equation}
v^{(1)}=\sqrt{1-\frac{N^{2}}{(E-\omega L)^{2}}(1+\frac{L^{2}}{g_{\phi \phi }}%
)}.  \label{v1}
\end{equation}

Then, introducing also the absolute value of the velocity $v$ according to%
\begin{equation}
v^{2}=\left[ v^{(1)}\right] ^{2}+\left[ v^{(2)}\right] ^{2}
\end{equation}%
one can find that%
\begin{equation}
E-\omega L=\frac{N}{\sqrt{1-v^{2}}}\text{,}  \label{e}
\end{equation}%
\begin{equation}
v^{2}=1-\left( \frac{N}{E-\omega L}\right) ^{2}.
\end{equation}

\section{Limiting transitions for relative velocity}

The energy $E_{c.m.}$ in the centre of mass frame of two colliding particles
can be defined as (see \cite{ban} and consequent papers)%
\begin{equation}
E_{c.m.}^{2}=-(p_{1}^{\mu }+p_{2}^{\mu })(p_{1\mu }+p_{2\mu
})=m_{1}^{2}+m_{2}^{2}-2m_{1}m_{2}u_{1}^{\mu }u_{2\mu }\text{.}
\end{equation}

Here, \thinspace $p_{i}^{\mu }=m_{i}u_{i}^{\mu }$ ($i=1,2$) is the
four-momentum of each particle, $m_{i}$ are their rest masses. By
definition, this is a scalar which can be calculated in any frame. It is
convenient to use a frame comoving with respect to one of colliding
particles (say, particle 2). If one uses tetrad representation, one can
exploit formulas known in a flat space-time. Then, the quantity of interest
is%
\begin{equation}
\gamma =-u_{1}^{\mu }u_{2\mu }=\frac{1}{\sqrt{1-w^{2}}}
\end{equation}%
where $w$ is, by definition, their relative velocity (which in this frame
coincides with the velocity of particle 1), $\gamma $ has the meaning of the
Lorentz factor.

The effect of unbound energies occurs if $w\rightarrow 1$, so $\gamma
\rightarrow \infty $.

Now, I remind some simple formulas from special relativity. If in the
laboratory frame particle 1 has the velocity $\vec{v}_{1}=v_{1}\vec{n}_{1}$
and particle 2 has the velocity $\vec{v}_{2}=v_{2}\vec{n}_{2}$, the value of
the relative velocity is equal to%
\begin{equation}
w^{2}=1-\frac{(1-v_{1}^{2})(1-v_{2}^{2})}{[1-v_{1}v_{2}(\vec{n}_{1}\vec{n}%
_{2})]^{2}}  \label{v}
\end{equation}

This formula can be found in textbooks (see. e.g., problem 1.3. in \cite{lt}%
). Now, we enumerate different limiting transitions for this quantity
relevant in our context.

a) $v_{1}\rightarrow 1,$ $v_{2}<1$, $(\vec{n}_{1}\vec{n}_{2})$ is arbitrary.

It is obvious from (\ref{v}) that in this case $w\rightarrow 1$ independent
of the quantity $(\vec{n}_{1}\vec{n}_{2}).$This corresponds to the well
known fact that the velocity of light $c$ is always equal to 1 (in
geometrical units) in any frame.

b) $v_{1}\rightarrow 1$, $v_{2}\rightarrow 1$ in such a way that $%
v_{i}=1-A_{i}\delta $ where $A_{i}$ ($i=1,2)$ are constants, $\delta \ll 1$.

b1). If $(\vec{n}_{1}\vec{n}_{2})\neq 1$, it is seen from (\ref{v}) that%
\begin{equation}
w^{2}\approx 1-\frac{4A_{1}A_{2}\delta ^{2}}{[1-(\vec{n}_{1}\vec{n}_{2})]^{2}%
}\text{,}
\end{equation}%
so we have $v\rightarrow 1$ again.

b2) If $(\vec{n}_{1}\vec{n}_{2})=1$, the situation changes radically. Then,%
\begin{equation}
w\approx \frac{\left\vert A_{1}-A_{2}\right\vert }{A_{1}+A_{2}}<1\text{.}
\label{a12}
\end{equation}

c) $v_{1}<1$, $v_{2}<1$, $(\vec{n}_{1}\vec{n}_{2})$ $\ $is arbitrary. Then,
it is obvious that $w<1$. By itself, this case is trivial. However, it plays
nontrivial role in the context under consideration (see below).

\section{Asymptotics near horizon}

Let us now look what happens to particles' velocities near the horizon. For
an usual particle, $E-\omega _{H}L\neq 0$ and it follows from (\ref{e}) that
in the horizon limit $N\rightarrow 0$, $v\rightarrow 1$. Apart from this, it
follows from (\ref{v3}), (\ref{v1}) that in this limit $v^{(3)}\rightarrow 0$%
, $v^{(1)}\rightarrow 1$. Therefore, the unit vector $\vec{n}$ is pointed
along the $l$ direction, so for any two such particles $(\vec{n}_{1}\vec{n}%
_{2})=1.$ 

However, for a critical particle, the situation is different. At first,
consider the extremal horizon. Then, near it, we have an expansion%
\begin{equation}
\omega =\omega _{H}-B_{1}N+B_{2}N^{2}+...  \label{ex}
\end{equation}%
For example, for the Kerr metric $B_{1}=M^{-1}$ where $B$ is the black hole
mass \cite{prd}. We obtain from (\ref{e}) that 
\begin{equation}
v^{2}=1-\frac{1}{L^{2}B_{1}^{2}}<1
\end{equation}

Apart from this, in the critical case the quantities $v^{(1)}$ and $v^{(3)}$
have the same order, so a particle hits the horizon at some nonzero angle
with respect to the normal direction in accordance with the remark made in 
\cite{ban}. Correspondingly, $(\vec{n}_{1}\vec{n}_{2})\neq 1$. Now, using
the above properties, we can enumerate different types of collisions near
the horizon.

\subsection{Collision between two usual particles}

This situation corresponds to case b2). Then, it follows form (\ref{a12})
that $w<1$, the Lorentz factor $\gamma $ is finite, so the effect of
infinite energies is absent.

\subsection{Collision between two critical particles}

This situation corresponds to case c). Then, we have that $w<1$, so the
effect under discussion is also absent.

\subsection{Collision between an usual (1) and critical (2) particles}

This type of collision falls into the class a) described above. As \ a
result, we have $w\rightarrow 1$, $\gamma \rightarrow \infty $ and the
effect of infinite acceleration is present. The fact that $v_{2}<1$ explains
why critical particle cannot reach the extremal horizon for a finite proper
time \cite{gp4}, \cite{prd}. Indeed, the proper distance is infinite, so the
proper time for a particle 2 having $v_{2}<1$ everywhere on its trajectory
is certainly infinite.

In the nonextremal case a near-critical particle cannot reach the horizon
since $\omega -\omega _{H}\sim N^{2}$ when $N\rightarrow 0$ \cite{04}, \cite%
{prd}, so the right hand side of (\ref{1}) cannot be positive. However, it
can approach the horizon as nearly as one likes. Let $E=\omega
_{H}L(1+\delta )$, $\delta \ll 1$. Then, we must keep $\delta $ such that $%
\delta \gtrsim N$ to ensure the positivity of $\dot{l}^{2}$ in (\ref{1}).
Let $\delta =AN(P)$ where $A$ is some finite coefficient, $P$ is the point
of collision. Then, $1-v^{2}=(\frac{N}{E-\omega _{H}L})^{2}\approx \frac{1}{%
(\omega _{H}LA)^{2}}\neq 0$. Thus, taking the point of collision closer and
closer to the horizon and simultaneously taking the energy closer and closer
to the critical value, we can gain $v<1$ and, thus, the effect of infinite
acceleration for the energy in the centre of mass for collision between an
usual and critical particles. However, this requires multiple scattering
since, say, for the Kerr metric, such a particle cannot come from infinity.
Apart from this, the collision should occur in a narrow strip near the
horizon (see \cite{gp4}, \cite{prd} for details).

\section{Charged static black holes}

All the above consideration applies also to charged static with minimum
changes. For simplicity, let us consider the spherically-symmetric black
holes. Then, equations of motion give us%
\begin{equation}
\dot{t}=u^{0}=\frac{E-\varphi q}{N^{2}}\text{,}  \label{oq}
\end{equation}%
\begin{equation}
\dot{\phi}=\frac{L}{g_{\phi \phi }},  \label{fq}
\end{equation}%
\begin{equation}
\dot{l}^{2}=\frac{(E-\varphi q)^{2}}{N^{2}}-1-\frac{L^{2}}{g_{\phi \phi }}
\label{1q}
\end{equation}%
where $\varphi $ is the electric potential with respect to infinity$.$The
tetrad basis can be obtained by putting $\omega =0$ in (\ref{h0}) - (\ref{h3}%
). One can find easily that%
\begin{equation}
v^{(3)}=\frac{LN}{\sqrt{g_{\phi \phi }}(E-\varphi q)}\text{,}
\end{equation}%
\begin{equation}
v^{(1)}=-\sqrt{1-\frac{N^{2}}{(E-\varphi q)^{2}}(1+\frac{L^{2}}{g_{\phi \phi
}})}
\end{equation}%
where $v^{(1)}<0$ since a particle is moving towards the horizon.

Now, instead of (\ref{e}), we have%
\begin{equation}
E-\varphi q=\frac{mN}{\sqrt{1-v^{2}}}  \label{eq}
\end{equation}%
where we restored explicitly in (\ref{eq}) the particle's rest mass $m$.

The condition of criticality is now $E-\varphi _{+}q=0$ where $\varphi _{+}$
is the potential of the black hole. Then, for the extremal case, $N\sim
r-r_{+}\sim \varphi _{+}-\varphi $ where $r$ is the standard curvature
coordinate, $r_{+}$ is the horizon radius. As a result, $v\neq 1$.

If $L\neq 0$, the previous consideration applies and we again obtain that
the effect under consideration is possible only when collision occurs
between an usual (1) and the critical (2) particles: $v_{1}\rightarrow 1$, $%
v_{2}<1$, so $w\rightarrow 1$, $\gamma \rightarrow \infty $. In doing so, an
usual particle hits the horizon perpendicularly whereas the critical one
does it at some incident angle, $(\vec{n}_{1}\vec{n}_{2})\neq 1$.

A\ new situation having no analog for rotating case, arises if $L=0$ \cite%
{jl}. Then, for all colliding particles $(\vec{n}_{1}\vec{n}_{2})=1$.
Nonetheless, the main conclusion about the effect produced by collision
between an usual and the critical particles is still valid.

In a similar way, for the nonextremal horizon the energy is finite but can
be made as large as one like if one uses near-critical particles with $%
E=\varphi _{+}q(1+\delta )$, where $\delta \sim N\ll 1$.

\section{Collision between massive and massless particles}

If one of particles is massless, the above explanation is not valid since
(i) there is no comoving frame for a massless particle, (ii) in any frame,
such \ a particle moves always with the velocity of light. Therefore,
kinematic explanation should be somewhat changed. For brevity, we call a
massive particle "electron" and a massless one "photon", although
consideration applies to any kinds of such particles.

We do not consider the case when both particles are massless. Classical
electrodynamics is linear theory, so interaction between photons could occur
due to weak quantum-electrodynamic effects only which we neglect.

We again consider the geodesic motion of particles in the equatorial plane $%
\theta =\frac{\pi }{2}$. For photons, the equations of motion have the form%
\begin{equation}
\frac{dt}{d\lambda }=k^{0}=\frac{\nu _{0}-\omega L_{2}}{N^{2}}\text{,}
\label{0p}
\end{equation}%
\begin{equation}
\frac{d\phi }{d\lambda }=\frac{L_{2}}{g_{\phi \phi }}+\frac{\omega (\nu
_{0}-\omega L_{2})}{N^{2}},
\end{equation}%
\begin{equation}
\left( \frac{dl}{d\lambda }\right) ^{2}=\frac{(\nu _{0}-CL_{2})^{2}}{N^{2}}-%
\frac{L_{2}^{2}}{g_{\phi \phi }},  \label{1m}
\end{equation}%
where $\nu _{0}=-k_{0}$, and $L_{2}=k_{\phi }$ are conserved frequency and
angular momentum, $k^{\mu }$ is the wave vector, $\lambda $ is the affine
parameter. The quantity $\nu _{0}$ has a meaning of frequency measured by a
remote observer at infinity where we assume that $\omega \rightarrow 0$, $%
N\rightarrow 1$.

Thus, the only difference in the form of equations between the massive (\ref%
{0}) - (\ref{1}) and massless cases reveals itself is in eqs. (\ref{1m}), (%
\ref{1}). We assume that $\frac{dt}{d\lambda }>0$, so that $\nu _{0}-\omega
L>0$ (motion forward in time), except, possibly on the horizon where we
admit the equality $\nu _{0}-\omega _{H}L_{1}=0$ (critical photon).

Now, the energy $E_{c.m.}$ in the centre of mass frame is given by the
expression%
\begin{equation}
E_{c.m.}^{2}=-(p^{\mu }+k^{\mu })^{2}
\end{equation}%
where the Planck constant 
h{\hskip-.2em}\llap{\protect\rule[1.1ex]{.325em}{.1ex}}{\hskip.2em}%
=1, $p^{\mu }=mu^{\mu }$, $m$ is the electron rest mass. Then, 
\begin{equation}
E_{c.m.}^{2}=m^{2}-2m(uk)\text{, }(uk)\equiv u^{\mu }l_{\mu }\text{.}
\label{en}
\end{equation}%
It follows from (\ref{0p}) - (\ref{1}) that%
\begin{equation}
-(uk)=\frac{X_{1}X_{2}-Z_{1}Z_{2}\text{ }}{N^{2}}-\frac{L_{1}L_{2}}{g_{\phi
\phi }}\text{,}  \label{uk}
\end{equation}%
where $\,X_{1}\equiv E_{1}-CL_{1}$, $X_{2}=\nu _{0}-\omega L_{2}$,%
\begin{equation}
Z_{i}=\sqrt{X_{i}^{2}-N^{2}b_{i}}\text{, }b_{1}=1+\frac{L_{i}^{2}}{g_{\phi
\phi }}\text{, }b_{2}=\frac{L_{2}^{2}}{g_{\phi \phi }},  \label{z1}
\end{equation}

When, repeating the straightforward calculations along the lines of \cite%
{prd} step by step, one can arrive at the conclusions that unbound grow of $%
E_{c.m.}^{2}$ is indeed possible if electron is critical, photon is usual or
vice versa. Meanwhile, it is more important to obtain qualitative
explanation of infinite grow of $E_{c.m.}^{2}$ without explicit calculation
of (\ref{uk}). To this end, we use again the ZAMO frame (\ref{h0}) - (\ref%
{h3}). We can obtain formulas for the photon. In contrast to (\ref{vi}), now
formulas for the $k^{\mu }$ do not contain denominator:%
\begin{equation}
k^{(i)}=k_{(i)}=k^{\mu }h_{\mu (i)}\text{, }k^{(0)}=k^{\mu }h_{\mu
}^{(0)}=-k^{\mu }h_{\mu (0)}\text{.}
\end{equation}%
This is due to the fact that instead of the proper time $\tau $ the
parameter $\lambda $ along the geodesics is used, the vector $k^{\mu }$
being light-like.

From equations of motion (\ref{0}) - (\ref{1})\ and formulas for tetrad
components, we have 
\begin{equation}
k^{(1)}=-\sqrt{\nu ^{2}-\frac{L^{2}}{g_{\phi \phi }}}\text{,}  \label{k1}
\end{equation}

\begin{equation}
k^{(3)}=\frac{L}{\sqrt{g_{\phi \phi }}},  \label{k3}
\end{equation}%
where we took sign "-" in (\ref{k1}) since we consider an ingoing photon.
The analog of eq. (\ref{e}) reads%
\begin{equation}
\nu =\frac{\nu _{0}-\omega N}{N}\text{.}  \label{w}
\end{equation}%
It can be also obtained writing the scalar $(uk)$ in two frames - the
original system (\ref{m}) and the ZAMO one.

Defining $k^{2}=\left[ k^{(1)}\right] ^{2}+\left[ k^{(2)}\right] ^{2}$, it
is seen that \ 
\begin{equation}
k^{2}=\frac{(\nu _{0}-\omega L)^{2}}{N^{2}}=\nu ^{2}
\end{equation}%
\begin{equation}
k^{(0)}=-k_{\mu }h_{(0)}^{\mu }=\frac{\nu _{0}-\omega L}{N}=\nu
\end{equation}%
as it should be for the lightlike vector since $k^{2}-\left( k^{(0)}\right)
^{2}=0$.

In the horizon limit $N\rightarrow 0,$ the component $v^{(3)}\rightarrow 0$, 
$v^{(1)}\rightarrow 1$ for an usual electron. Therefore, the unit vector $%
\vec{n}_{1}=\frac{\vec{v}}{v}$ is pointed along $l$ direction,
perpendicularly to the horizon. For the critical particle this is not so 
\cite{ban} since $v^{(1)}\sim v^{(3)}$ have the same order. The similar
properties hold in the case of a photon for the vector $\vec{n}_{2}=\frac{%
\vec{k}}{k}$. Thus, in the horizon limit $(\vec{n}_{1}\vec{n}_{2})=1$ when
both particles are usual and $(\vec{n}_{1}\vec{n}_{2})\neq 1$ in other cases.

\subsection{Different types of collisions}

Now, we consider separately different cases depending on which particle (if
any) is critical.

\subsubsection{Case 1: electron is critical, photon is usual}

Let us pass to the frame which is comoving with respect to the electron.
Then, the frequency $\nu ^{\prime }$ measured in this frame is related to
the frequency $\nu $ in the ZAMO frame by the standard relativistic formula%
\begin{equation}
\nu ^{\prime }=\gamma (\nu -\vec{k}\vec{v})=\nu \gamma \lbrack 1-v(\vec{n}%
_{1}\vec{n}_{2})]\text{. }  \label{om}
\end{equation}

For a critical particle, as is explained above, $v\neq 1$, so the Lorentz
factor $\gamma $ is finite. The scalar product $(\vec{n}_{1}\vec{n}_{2})\neq
1$, the quantity $\nu ^{\prime }$ has the order $\nu .$ But, as a photon is
usual, $\nu \rightarrow \infty $. Thus, $\nu ^{\prime }\rightarrow \infty $
as well, so the effect reveals itself.

The resulting effect can be interpreted as a consequence of two factors. On
one hand, there is an infinite blue shift of radiation due to strong
gravitating field near a black hole. From the other hand, there is red shift
due to the Doppler effect since in the laboratory frame a receiver of
radiation is moving apart from a photon (both $v^{(1)}<0$ and $k^{(1)}<0$).
It turned out that in the case under discussion the first factor is infinite
whereas the second one is finite, so the net outcome is due to blue shift.

\subsubsection{Case 2: electron is usual, photon is critical}

As the photon is critical,\thinspace\ $\nu $ is finite. But, as the electron
is usual, $v\rightarrow 1$, $\gamma \rightarrow \infty $. The quantity $(%
\vec{n}_{1}\vec{n}_{2})\neq 1$. Thus, as a result, $\nu ^{\prime
}\rightarrow \infty $ and we again obtain the effect under discussion.

Interpretation again involves the Doppler effect but the concrete details
change. Let in a flat space-time a photon with the frequency $\nu $
propagate in the laboratory frame and some observer moves with the velocity $%
v$ with respect to this frame. Then, in in its own frame, the observer
measures the frequency of the process which is equal to $\nu ^{\prime }$. In
the case under discussion, $(\vec{n}_{1}\vec{n}_{2})\neq 1$. For simplicity,
we can take $(\vec{n}_{1}\vec{n}_{2})=0$. Then, the frequency measured in
the frame of a receiver $\nu ^{\prime }=\nu \gamma >\nu $ due to the
transverse Doppler effect. In the limit $v\rightarrow 1$, the Lorentz factor 
$\gamma \rightarrow 1$ and the frequency $\nu ^{\prime }\rightarrow \infty $%
. In other words, even despite a moderate gravitational blue shift that
resulted in a finite $\nu $, the net outcome is infinite due to the Doppler
effect.

\subsubsection{Case 3: both particles are critical}

Then, $(\vec{n}_{1}\vec{n}_{2})=1$ but $v<1,\nu $ is finite. It follows from
(\ref{om}) that $\nu ^{\prime }$ is also finite, so there is no effect under
discussion. In other words, both factors - gravitational blue shifting and
the Doppler effect are restricted and cannot give rises to infinite energies.

\subsubsection{Case 4: both particles are usual}

Here, an accurate estimate of different terms in the horizon limit is
required. In the limit $N\rightarrow 0$ the quantities $\gamma \sim \frac{1}{%
N}$, $\nu \sim \frac{1}{N}$ as it is seen from (\ref{e}), (\ref{w}). It
follows also from (\ref{v1}), (\ref{v3}), (\ref{k1}), (\ref{k3}) that

\begin{equation}
1-(\vec{n}_{1}\vec{n}_{2})\sim N^{2}\text{.}
\end{equation}

As a result, the factors $N^{2}$ in the numerator and denominator compensate
each other, $\nu ^{\prime }$ remains finite, the effect of infinite
acceleration is absent. One can say that the effect of infinite red shift
due to Doppler effect for a receiver moving apart from the photon is
completely compensated by an infinite blue shifting the photon frequency.

\section{Conclusion}

We gave a simple and general explanation of the effect of infinite energy in
the centre of mass of particles colliding near the horizon of a black hole.
It is based on kinematics of particles in a flat space-time plus properties
of the horizon. It is given for massive and massless particles separately.

For massive ones, it is essential that in the ZAMO frame (or static one in
the case of static charged black holes) (i) usual particles have the
velocities approaching the speed of light near the horizon, (ii) for a
special class of critical particles this limit differs from the speed of
light. Then, collision between an usual and critical particles produces the
effect under discussion. Thus, critical particles play a distinguished role
in the kinematics of the process. These particles have also some other
properties that distinct them from usual ones: they hit the horizon
nonperpendicularly (in the case of rotating black holes), the proper time
required to reach the extremal horizon is infinite. The distinction between
usual and critical particles is also seen from the simple formulas for the
particle's energy in the stationary gravitational field (\ref{e}), (\ref{eq}%
) which show what happens to the velocity when a particle approaches the
horizon.

For massless particles, we showed that, again, the distinguished role is
played by critical particles although their definition and properties are
somewhat different. Now, interpretation in terms of the velocity is not
valid. Instead, it is done in terms of the frequency: for photons their
frequency in the same frame remains finite notwithstanding the vanishing the
lapse function near the horizon. The crucial point is that the BSW effect is
possible only for the case when one and only one of colliding particles is
critical. The role of critical particles gave rise to natural classification
taking into account two factors - gravitational blue shift (GB) and the
Doppler effect (DE). Namely, we have four cases: 1) critical electron, usual
photon: infinite GB, finite DE, $E_{c.m.}$ is infinite, 2) critical photon,
usual electron: finite GB, infinite DE, $E_{c.m.}$ is infinite, 3) both
particles are critical: finite GB, finite DE, $E_{c.m.}$ is finite, 4) both
particles are usual: infinite GB, infinite DE, $E_{c.m.}$ is finite due to
their compensation. The corresponding results can be used for investigation
of the Compton effect near black holes. Meanwhile, the possibility of
infinite $E_{c.m.}$ means that, apart from mutual scattering of electrons
and photons, qualitatively new reactions can occur with creation of new
kinds of high energy particles.

The above consideration is based on test particle approximation, with
backreaction, gravitation and electromagnetic radiation neglected. Whether
and how these results can be changed if these factors are taken into
account, remains an interesting task for further studies.

\end{document}